\documentclass[12pt,prd,nopacs,tightenlines,nofootinbib]{revtex4}
\usepackage{bm}
\usepackage{graphics}
\usepackage{rotating}
\usepackage{epsfig}
\usepackage{slashed}
\usepackage{amsmath}

\usepackage{multirow}
\begin{document}
\title{Rare radiative $\Xi^{-}_{b}\to \Xi^-\gamma$ decay in the
  relativistic quark model}
\author{A.O. Davydov $^1$}
\author{R. N. Faustov $^2$}
\author{V. O. Galkin $^2$}
\affiliation{$^1$ Faculty of Physics, M.V. Lomonosov Moscow State University,  119991 Moscow, Russia\\
  $^2$ Federal Research Center ``Computer Science and Control'', Russian Academy of Sciences,  Vavilov Street 40, 119333 Moscow, Russia}

\begin{abstract}
Recently the LHCb collaboration put the upper limit on the
$\Xi^{-}_{b} \rightarrow \Xi^{-}\gamma$ decay branching ratio
$Br(\Xi^{-}_{b} \rightarrow \Xi^{-}\gamma)< 1.3\times 10^{-4}$. The
measured value is below the light-cone sum rule  prediction. In this
paper the rare radiative decay of the $\Xi^{-}_{b}$ baryon is studied
in the framework of the relativistic quark model based on the
quasipotential approach and QCD. The decay form factors are calculated
with the comprehensive account of the relativistic effects.  The
obtained result for the branching ratio is found to be below the upper
limit set by LHCb and is consistent with theoretical predictions based
on the SU(3) flavor-symmetry, light-front quark model and light-cone QCD sum rules in full theory within  theoretical uncertainties.

\end{abstract}

\maketitle
\section{ Introduction}

In the standard model (SM) the exclusive rare weak decays of hadrons
governed by the $b \rightarrow s$ quark transitions proceed through
the flavor-changing neutral currents (FCNC). Thus such processes are
forbidden at the tree level in SM. The leading contribution comes from
the one-loop, so-called penguin diagrams. As a result, such a decay
channel is strongly suppressed, which complicates its experimental
search. The rare radiative decay $\Xi^{-}_{b}(bsd)\to \Xi^-(ssd)\gamma$ has not
yet been observed experimentally. However, in 2021
the LHCb collaboration at Large Hadron Collider  (CERN) set an
upper experimental limit on its branching ratio
\cite{Experimental}. Note that the branching ratio of the similar rare
radiative decay $\Lambda_{b}(bud) \rightarrow \Lambda(sud) \gamma$ was
measured by the LHCb Collaboration in 2019 \cite{Lambda}.

Theoretically the rare radiative decay of the $\Xi^{-}_{b}$ baryon has
been studied with different approaches. The prediction based on  the
$SU(3)$-flavor symmetry, which relates the decay branching fraction of the $\Xi_b$ baryon
to the measured branching fraction of the $\Lambda_b$ baryon,  gives
the value consistent with the
upper experimental limit \cite{SU(3)}. The light-front quark model
\cite{lfqm} and  light-cone QCD sum rules within
full theory predict the value satisfying this upper limit too
\cite{Rare}, while the computation using light-cone sum rules shows a
significant tension with the experimental value
\cite{2011}. Therefore a more detailed theoretical investigation of
this decay is required.

In this paper we comprehensively investigate the rare radiative decays $\Xi^{-}_{b} \rightarrow \Xi^{-}\gamma$ in the framework of the relativistic quark-diquark model based on the quasipotential 
approach.  All relativistic effects, including the wave function
transformations from the rest to the moving reference frame and
contributions of the intermediate negative-energy states, are
systematically taken into account. Using baryon wave functions,
found in the previous studies of the baryon spectroscopy, we
calculate the form factors parameterizing the baryon decay matrix element. The
obtained form factors of the $\Xi^{-}_{b}$ baryon transitions are used
for the prediction of the $\Xi^{-}_{b} \rightarrow \Xi^{-}\gamma$
decay branching ratio. Our result is consistent with the values from
Refs.~\cite{SU(3),lfqm,Rare} within theoretical uncertainties and
shows a small deviation in central values. Note that all these
predictions are lower than experimental limit.

There are good chances that this decay will be soon observed by the
LHCb Collaboration. So it will be possible to compare results obtained with different methods and decide which method gives a more precise prediction.

\section{Effective Hamiltonian}

The effective Hamiltonian for the rare $b \rightarrow s$ transitions is given by \cite{Buras}:
\begin{eqnarray}\label{effhamilt}
     {\cal H}^{\rm eff} &=&
                        -\frac{G_{F}}{\sqrt{2}}V_{tb}V^{*}_{ts}\Bigg[\sum_{i=1}^{6} C_{i}(\mu)Q_{i}(\mu) + C_{7\gamma}(\mu)Q_{7\gamma}(\mu) + C_{8G}(\mu)Q_{8G}(\mu)\Bigg]  \cr &\equiv& -\frac{G_{F}}{\sqrt{2}}V_{tb}V^{*}_{ts}\vec{Q^{T}}(\mu)\vec{C}(\mu)
\end{eqnarray}
   
Here $C_{i}(\mu)$ are the Wilson coefficients, $G_{F}$ is the Fermi
coupling constant, $V_{tb}$ and $V^{*}_{ts}$ are the 
Cabibbo-Kobayashi-Maskawa (CKM) matrix elements, $Q_{i}(\mu)$ are the local operators given by
\begin{eqnarray*}
    Q_{1}&=&(\bar{s}_{\alpha}c_{\beta})_{V-A}(\bar{c}_{\beta}b_{\alpha})_{V-A},\cr\cr
    Q_{2}&=&(\bar{s}_{\alpha}c_{\alpha})_{V-A}(\bar{c}_{\beta}b_{\beta})_{V-A},
\cr\cr
    Q_{3}&=&(\bar{s}_{\alpha}b_{\alpha})_{V-A}\sum_{q}(\bar{q}_{\beta}q_{\beta})_{V-A},\cr
    Q_{4}&=&(\bar{s}_{\beta}b_{\alpha})_{V-A}\sum_{q}(\bar{q}_{\alpha}q_{\beta})_{V-A},\cr
    Q_{5}&=&(\bar{s}_{\alpha}b_{\alpha})_{V-A}\sum_{q}(\bar{q}_{\beta}q_{\beta})_{V+A},\cr
    Q_{6}&=&(\bar{s}_{\beta}b_{\alpha})_{V-A}\sum_{q}(\bar{q}_{\alpha}q_{\beta})_{V+A},\cr
    Q_{7\gamma} &=& \frac{e}{4\pi^2}\bar{s}_{\alpha}\sigma^{\mu\nu}(m_{b}R+m_{s}L)b_{\alpha}F_{\mu\nu},\cr\cr
Q_{8G}& =& \frac{g_{s}}{4\pi^2}\bar{s}_{\alpha}\sigma^{\mu\nu}(m_{b}R+m_{s}L)T^{a}_{\alpha\beta}b_{\beta}G^{a}_{\mu\nu},
\end{eqnarray*}
where $(\bar{q}_{\alpha}q_{\beta})_{V\pm A} = \bar{q}_{\alpha}\gamma_{\mu}(1\pm\gamma_{5})q_{\beta}$, $\vec{Q^{T}} = $ $(Q_{1},Q_{2},...,Q_{8G})$, $\vec{C^{T}} = (C_{1},C_{2},...,C_{8G})$, $\alpha$ and $\beta$ are color indices, $R=\frac{1+\gamma_{5}}{2}$ and $L=\frac{1-\gamma_{5}}{2}$, $\gamma_{\mu}$ and $\gamma_{5}$ are the Dirac matrices, $e$ and $g$ are electromagnetic and strong coupling constants, respectively. $F_{\mu\nu}$ is the electromagnetic field strength tensor, which in the case of the plane electromagnetic wave is given by
\[
    F_{\mu\nu} = -i(\epsilon_{\mu}q_{\nu} - \epsilon_{\nu}q_{\mu})e^{iqx},
\]
where $\epsilon_{\mu}$ is the polarization four-vector, $q_{\mu}$ is
the four-momentum vector, and $\sigma_{\mu\nu}$ is a commutator of the
Dirac matrices
\[
    \sigma_{\mu\nu} = \frac{i}{2} \big[\gamma_{\mu}\gamma_{\nu}-\gamma_{\nu}\gamma_{\mu}\big].
\]

For the $b\rightarrow s\gamma$ transition the main contribution comes from the magnetic penguin operator $Q_{7\gamma}$, thus the effective Hamiltonian takes the form

\begin{equation}
    {\cal H}^{\rm eff} =
    -\frac{G_{F}\, e}{4\pi^2\sqrt{2}}V_{tb}V^{*}_{ts}C^{\rm eff}_{7\gamma}(m_{b})\bar{s}\sigma_{\mu\nu}\bigg[m_{b}R+m_{s}L\bigg]bF^{\mu\nu}.
\end{equation}

We first calculate the value of the relevant effective Wilson coefficient $C^{\rm eff}_{7\gamma}$ in the leading order. To achieve this goal we need to solve the system of renormalization group equations
\begin{equation}\label{effective-equiv}
    \frac{d\vec{C}^{(0)\rm eff}(\mu)}{d\ln{\mu}} =
    \frac{\alpha_{s}}{4\pi}\big(\hat{\gamma}^{(0)\rm eff}\big)^{T}\vec{C}^{(0)\rm
      eff}(\mu),
\end{equation}
where index $(0)$ stands for leading order, $\alpha_{s}(\mu)$ is the running strong coupling constant
\begin{equation}\label{constant}
    \alpha_{s}(\mu) = \frac{\alpha_{s}(M_{Z})}{1-\frac{\beta_{0}}{2\pi}\alpha_{s}(M_{Z})\ln{\frac{M_{Z}}{\mu}}},
\end{equation}
\begin{equation}
    \beta_{0}=\frac{11N_{c}-2n_f}{3} 
\end{equation}
with the number of colors $N_{c}=3$, and  the number of quark flavors
$n_f=5$. We take the current world averaged value \cite{PDG} of
\begin{equation}
    \alpha_{s}(M_{Z}) = 0.1179 \pm 0.0010.
\end{equation}
The effective anomalous dimension matrix $\hat{\gamma}^{(0)\rm eff}$
is given by  \cite{Buras}, \cite{Ciuchini}
\begin{equation}
    \hat{\gamma}^{(0)\rm eff} = 
    \begin{pmatrix}
  -2 & 6 & 0 & 0 & 0 & 0 & 0 & 3 \\ \\
  6 & -2 & -\frac29 & \frac23 & -\frac29 & \frac23 & \frac{416}{81} & \frac{70}{27} \\ \\
   0 & 0 & -\frac{22}{9} & \frac{22}{3} & -\frac{4}{9} & \frac{4}{3} & -\frac{464}{81} & \frac{545}{27} \\ \\
   0 & 0 & \frac{44}{9} & \frac43 & -\frac{10}{9} & \frac{10}{3} & \frac{136}{81} & \frac{512}{27} \\ \\
   0 & 0 & 0 & 0 & 2 & -6 & \frac{32}{9} & -\frac{59}{3} \\ \\
   0 & 0 & -\frac{10}{9} & \frac{10}{3} & -\frac{10}{9} & -\frac{38}{3} & -\frac{296}{81} & -\frac{703}{27} \\\\ 
   0 & 0 & 0 & 0 & 0 & 0 & \frac{32}{3} & 0 \\\\ 
   0 & 0 & 0 & 0 & 0 & 0 & -\frac{32}{9} & \frac{28}{3}
\end{pmatrix}
\end{equation}
The initial conditions are as follows \cite{Buras}
\begin{equation}\label{ic}
    \begin{cases}
     C^{0}_{1}(M_{W}) = C^{0}_{3}(M_{W}) = C^{0}_{4}(M_{W}) = C^{0}_{5}(M_{W}) = C^{0}_{6}(M_{W}) = 0, \\
     C^{0}_{2}(M_{W}) = 1, \\
     C^{(0)\rm eff}_{7\gamma}(M_{W}) = \frac{3x^3-2x^2}{4(x-1)^4}\ln{x} + \frac{-8x^3-5x^2+7x}{24(x-1)^3} \approx -0.194 , \\
     C^{(0)\rm eff}_{8G}(M_{W}) = -\frac{3x^2}{4(x-1)^4}\ln{x} + \frac{-x^3+5x^2+2x}{8(x-1)^3} \approx -0.097,
    \end{cases}
\end{equation}
where $x\equiv \frac{m^2_{t}}{(M^2_{W})} \approx 4.62$.

Solving Eq.~(\ref{effective-equiv}) with the initial conditions
(\ref{ic}) we obtain the expression for the effective Wilson coefficient
$C^{(0)\rm eff}_{7\gamma}$
\begin{equation}
    C^{(0)\rm eff}_{7\gamma}(\mu) = \eta^{\frac{16}{23}}C^{(0)}_{7\gamma}(M_{W}) + \frac{8}{3}\bigg(\eta^{\frac{14}{23}}-\eta^{\frac{16}{23}}\bigg)C^{(0)}_{8G}(M_{W}) + C^{(0)}_{2}(M_{W})\sum^{8}_{j=1}K_{j}\eta^{c_{j}},
\end{equation}
where
\begin{equation}
    K_{j} = \bigg(-\frac{3}{7}, -\frac{1}{14}, -0.6494, -0.0380, -0.0185, -0.0057,  2.2996, -1.0880 \bigg),
  \end{equation}
  \begin{equation}
    c_{j} = \left(\frac{6}{23}, -\frac{12}{23}, 0.4086, -0.4230, -0.8994, 0.1456,  \frac{14}{23}, \frac{16}{23} \right),
\end{equation}
and
\begin{equation*}
    \eta \equiv \frac{\alpha_{s}(\mu_{M_{W}})}{\alpha_{s}(\mu)}.
\end{equation*}

Substituting numerical values we get the following result for the
effective Wilson coefficient $C^{(0)\rm eff}_{7\gamma}$ at $\mu=m_{b}$
\begin{equation}
    C^{(0)\rm eff}_{7\gamma}(m_{b}) = 0.674C^{(0)}_{7\gamma}(M_{W}) + 0.091C^{(0)}_{8G}(M_{W}) - 0.170C^{(0)}_{2}(M_{W}) = -0.310. 
\end{equation}

\section{Relativistic quark model}

Now we  calculate the matrix element of the effective Hamiltonian
${\cal H}^{\rm eff}$ between the initial and final states
\begin{equation}
     M = \big<\Xi^{-}\gamma\big|{\cal H}^{\rm eff}|\Xi^{-}_{b}\big>.
\end{equation}

Note that in the absence of the QCD corrections we can make the
following replacement in ${\cal H}^{\rm eff}$:
$-\sigma^{\mu\nu}F_{\mu\nu} \rightarrow 2i\sigma^{\mu\nu}\epsilon_\mu
q_{\nu}$. As a result, the operator $Q_{7\gamma}$ reduces to
$-2i\frac{e}{4\pi^2}\bar{s}_{\alpha}\sigma^{\mu\nu}\epsilon_\mu q_{\nu}(m_{b}R+m_{s}L)b_{\alpha}$ \cite{Buras}.
Thus  to find $M$ we need to calculate the following matrix elements between baryon states
\begin{equation}\label{vec}
     \big<\Xi^{-}(P)\big|\bar{s}\sigma^{\mu\nu}q_{\nu}b|\Xi^{-}_{b}(Q)\big>,
\end{equation}
and
\begin{equation}\label{ax}
     \big<\Xi^{-}(P)\big|\bar{s}\sigma^{\mu\nu}q_{\nu}\gamma_{5}b|\Xi^{-}_{b}(Q)\big>.
   \end{equation}

   The matrix elements~(\ref{vec}) and~(\ref{ax}) can be parameterized by
the following set of form factors \cite{Galkin}
\begin{equation}\label{vector}
    \big<\Xi^{-}|\bar{s}i\sigma^{\mu\nu}q_{\nu}b|\Xi^{-}_{b}\big> = \bar{u}_{\Xi^{-}}(p_{\Xi},s)\bigg[\frac{f^{T}_{1}(q^2)}{m_{\Xi^{-}_{b}}}(\gamma^{\mu}q^2-\slashed{q}q^{\mu})-f^{T}_{2}(q^2)i\sigma^{\mu\nu}q_{\nu}\bigg]u_{\Xi^{-}_{b}}(p_{\Xi_{b}},s^{'}),
\end{equation}
\begin{equation}\label{axial}
    \big<\Xi^{-}|\bar{s}i\sigma^{\mu\nu}q_{\nu}\gamma_{5}b|\Xi^{-}_{b}\big> = \bar{u}_{\Xi^{-}}(p_{\Xi},s)\bigg[\frac{g^{T}_{1}(q^2)}{m_{\Xi^{-}_{b}}}(\gamma^{\mu}q^2-\slashed{q}q^{\mu})-g^{T}_{2}(q^2)i\sigma^{\mu\nu}q_{\nu}\bigg] \gamma_{5}u_{\Xi^{-}_{b}}(p_{\Xi_{b}},s^{'}).
\end{equation}
Since we investigate the decay with the emission of the real photon, we only need the values of the form factors $f^{T}_{2}(q^2)$ and $g^{T}_{2}(q^2)$ at $q^2=0$.

Summing the expressions~(\ref{vector}) and~(\ref{axial}) we get the
following parameterization for the matrix element of the effective
Hamiltonian between baryon states
\begin{multline}
    \big<\Xi^{-}\gamma|{\cal H}^{\rm eff}|\Xi^{-}_{b}\big>  \\ =
    \frac{G_{F}m_{b}e}{4\pi^2\sqrt{2}}V_{tb}V^{*}_{ts}C^{(0)\rm
      eff}_{7\gamma}(m_{b})\bar{u}_{\Xi^{-}}(p_{\Xi},s)i\sigma^{\mu\nu}\epsilon_\mu
    q_{\nu}\bigg(g_{V}f^{T}_{2}(0)+\gamma_{5}g_{A}g^{T}_{2}(0)\bigg)   u_{\Xi^{-}_{b}}(p_{\Xi_{b}},s^{'}),
\end{multline}
where $g_{V} = 1 + \frac{m_{s}}{m_{b}}$ and $g_{A} = 1 - \frac{m_{s}}{m_{b}}$. 
   
For the evaluation of the form factors $f_2^T(0)$ and $g_2^T(0)$ we employ the relativistic quark-diquark
model. In the quasipotential approach the matrix element of a local current $J_{\mu}$ is given by \cite{Galkin}

\begin{equation}
    \big<\Xi^{-}(P)\big|J_{\mu}|\Xi^{-}_{b}(Q)\big> = \int
    \frac{d^{3}pd^{3}q}{(2\pi)^6}\bar{\Psi}_{\Xi^{-}_{\bf
        P}}(\textsc{{\bf p}})\Gamma_{\mu}(\textsc{{\bf
        p}},\textsc{{\bf q}})\Psi_{\Xi^{-}_{b{\bf Q}}}(\textsc{{\bf q}}),
\end{equation}
where $P$ and $Q$ are momenta of the final and initial baryons, respectively, and $\Gamma_{\mu}(\textsc{{\bf p}},\textsc{{\bf q}})$ is the two-particle vertex function.
In our case $\Gamma_{\mu}(\textsc{{\bf p}},\textsc{{\bf q}}) = \Gamma^{(1)}_{\mu}(\textsc{{\bf p}},\textsc{{\bf q}}) + \Gamma^{(2)}_{\mu}(\textsc{{\bf p}},\textsc{{\bf q}})$, where:

\begin{equation*}
    \Gamma^{(1)}_{\mu}(\textsc{{\bf p}},\textsc{{\bf q}}) =
    \psi^{*}_{d}(p_{d})\bar u_{s}(p_{s})\gamma_{\mu}(1-\gamma_{5})u_{b}(q_{b})\psi_{d}(q_{d})(2\pi)^3\delta(\textsc{{\bf p}}_{d}-\textsc{{\bf q}}_{d})
\end{equation*}
is the vertex function corresponding to the impulse approximation
diagram (see Fig.~1 from \cite{Galkin} for the analogous process $\Lambda_{b} \rightarrow \Lambda\gamma$), 
while 
\begin{multline*}
    \Gamma^{(2)}_{\mu}(\textsc{{\bf p}},\textsc{{\bf q}}) =
    \psi^{*}_{d}(p_{d})\bar
    u_{s}(p_{s})\bigg[\gamma_{\mu}(1-\gamma_{5})\frac{\Lambda^{(-)}_{b}(k)}{\epsilon_{b}(k)+\epsilon_{b}(p_{s})}\gamma^{0}{\cal
      V}(\textsc{{\bf
        p}}_{d}-\textsc{{\bf q}}_{d})  + \\ + {\cal V}(\textsc{{\bf p}}_{d}-\textsc{{\bf q}}_{d})\frac{\Lambda^{(-)}_{s}(k^{'})}{\epsilon_{s}(k^{'})+\epsilon_{s}(q_{b})}\gamma^{0}\gamma_{\mu}(1-\gamma_{5})\bigg]u_{b}(q_{b})\psi_{d}(q_{d})
\end{multline*}
is the vertex function corresponding the diagrams (see Fig.~2 from
\cite{Galkin} for the analogous process $\Lambda_{b} \rightarrow \Lambda\gamma$) with the intermediate negative-energy states which are the consequence of the projection onto the positive-energy states in the quasipotensial approach. 
Here $\psi_{d}(p)$ is the diquark wave function, ${\cal V}(\textsc{{\bf p}})$ is the quark-diquark interaction quasipotential, $\bf k$ = $\bf p_{s}-\bm \Delta$, 
$\bf k^{'} = \bf q_{b}+\bm \Delta$,
$\bm \Delta = \bf P-\bf Q$, $\epsilon(p)=\sqrt{m^2+\textsc{{\bf p}}^2}$, and
\begin{equation*}
    \Lambda^{(-)}(p)=\frac{\epsilon(p)-(m\gamma^{0}+\gamma^{0}({\bm{\gamma}} \bf p))}{2\epsilon(p)},
\end{equation*}
and $u_q(p)$  are the Dirac bispinors, $m_{q}$ and $m_{d}$
 are the quark and diquark masses, respectively.

 The baryon wave functions $\Psi_{\Xi^{-}_{b\bf Q}}({\bf q})$ and
 $\Psi_{\Xi^{-}_{\bf P}}({\bf p})$ are  projected onto the positive-energy
 states of quarks and boosted to the moving reference frame. Indeed, in the
 rest frame of the initial baryon $\Xi^{-}_b$ the final baryon is moving with the recoil momentum ${ \bf P}$.
Thus we have to boost the wave function of the final baryon $\Xi^{-}$ to the moving reference frame
\begin{equation}
    \Psi_{{\Xi^{-}_{\bf P}}}(\textsc{{\bf p}}) =
    D^{1/2}_{q}(R^{W}_{L_{P}})D_{d}(R^{W}_{L_{P}})\Psi_{\Xi^{-}{\bf 0}}({{\bf p}}),
\end{equation}
where $\Psi_{\Xi^{-}{\bf 0}}({{\bf p}})\equiv \Psi_{\Xi^{-}}({\bf p})$ is the baryon wave function in
the rest frame, $R^{W}$ is the Wigner rotation, $L_{P}$ is the Lorentz
boost from the baryon rest frame to a moving one with the momentum
{\bf P}, and
$D^{1/2}_{q}(R^{W})$ is the rotation matrix of the quark spin, while
the rotation matrix for the scalar diquark  $D_{d}(R^{W})=1$.

 The baryon $B=\Xi_b$ or $\Xi$ wave functions in the rest frame satisfy the relativistic quasipotential equation of the Schrödinger type
\begin{equation}\label{quasiequv}
    \bigg(\frac{b^{2}(M)}{2\mu_{R}(M)} - \frac{{{\bf p}}^2}{2\mu_{R}(M)}\bigg)\Psi_B({{\bf p}}) = \int\frac{d^{3}q}{(2\pi)^3}V({{\bf p}}, {{\bf q}}, M)\Psi_B({{\bf q}}),
\end{equation}
where
\begin{equation*}
    \mu_{R}(M) \equiv \frac{M^4_{B} - (m^2_{q}-m^2_{d})^2}{4M^3_{B}}
\end{equation*}
is the relativistic reduced mass, and

\begin{equation*}
    b^{2}(M) = \frac{(M^2_{B} - (m_{q}+m_{d})^2)(M^2_{B} - (m_{q}-m_{d})^2)}{4M^2_{B}}
\end{equation*}
is the relativistic center-of-mass system relative momentum squared on the mass
shell. The quark-diquark interaction potential $V({\bf p,q},M)$ is
constructed from the off-mass-shell scattering amplitude  projected on
the positive energy states. It 
includes all spin-dependent and spin-independent relativistic
contributions. Its explicit form can be found
Ref.~\cite{Galkpotensial}.

Explicit expressions for the form factors are given in
Ref.~\cite{Galkin}. Substituting the baryon wave functions which were
obtained in the baryon mass calculations \cite{Galkcurrent} we obtain
the following values of form factors at $q^2=0$
\begin{equation}
    f^{T}_{2}(0) =  g^{T}_{2}(0) = -0.144.
\end{equation}
We estimate the uncertainties of the calculated values of the form
factors to be less than 5\%.

\section{Rare radiative decay rate}

\begin{table}
\caption{Values of the physical constants.}
\label{tabular:1}
\begin{center}
\begin{tabular}{c*{1}{c}}
\hline
Quantity & Numerical value \\
\hline
$G_{F}$ & $1.166\times 10^{-5}$ GeV$^{-2}$ \\ 
$\alpha_{em}(M_{W})$ & 1/128 \\ 
$|V_{ts}|$ & $(38.8\pm 1.1)\times 10^{-3}$ \\ 
$|V_{tb}|$ & $1.013\pm 0.030$ \\ 
$m_{b}$(pole) & $(4.78\pm 0.06)$ GeV \\ 
$m_{s}$ & $93^{+11}_{-5}$ MeV \\ 
$m_{\Xi^{-}_{b}}$ & $(5797.0 \pm 0.6)$ MeV \\ 
$m_{\Xi^{-}}$ & $(1321.71 \pm 0.07)$ MeV \\ 
$M_{W}$ & $(80.379 \pm 0.012)$ GeV \\
$M_{Z}$ & $(91.1876 \pm 0.0021)$ GeV \\
$m_{t}$ & $(172.76 \pm 0.30)$ GeV \\
$\tau_{\Xi_{b}}$ & $(1.572 \pm 0.040)\times 10^{-12} s$\\ 
$\hbar$ & $6.582\times 10^{-22}$ MeV $s$ \\ \hline
\end{tabular}
\end{center}
\end{table}

The exclusive rare radiative decay rate $\Xi^{-}_{b} \rightarrow \Xi^{-} \gamma$ for the
emission of a real photon $q^2=0$ is given by
\begin{equation}
    \Gamma
    =\frac{G^2_{F}\alpha_{em}}{64\pi^4}|V_{tb}V^{*}_{ts}|^2m^{2}_{b}|C^{(0)\rm eff}_{7\gamma}(m_b)|^2(g^2_{V}|f_{2}^{T}(0)|^2+g^2_{A}|g_{2}^{T}(0)|^2)\Bigg(\frac{m^2_{\Xi^{-}_{b}}-m^2_{\Xi^{-}}}{m_{\Xi^{-}_{b}}}\Bigg)^3,
\end{equation}
where $\alpha_{em} \equiv \frac{e^2}{4\pi}$ is the electromagnetic coupling constant.

\begin{table}
\caption{The calculated values of the effective Wilson coefficient and
  form factors at $q^2=0$.}
\label{tabular:2}
\begin{center}
    \begin{tabular}{cc}
      \hline
      Quantity&Numerical value\\ \hline
$C^{(0) \rm eff}_{7}(m_{b})$ &  $-0.310$ \\ 
$f_2^{T}(0)=g_2^{T}(0)$ & $-0.144$ \\ \hline
    \end{tabular}   
\end{center}
\end{table}

Substituting the values of the physical constants summarized in
Table~\ref{tabular:1} \cite{PDG} and the calculated values of the form factors given in Table~\ref{tabular:2}
we get the prediction for the branching fraction
\begin{equation}
    Br(\Xi^{-}_{b} \rightarrow \Xi^{-}\gamma)=(0.95 \pm 0.15)\times 10^{-5}.
\end{equation}

\begin{table}
  \caption{Comparison of the theoretical predictions with experimental
    upper limit for the branching
    fraction of the $\Xi_b^-\to \Xi^-\gamma$ decay.}
\label{tabular:timesandtenses}
\begin{center}
    \begin{tabular}{c*{1}{c}}
\hline
 Reference & Predicted value \\ \hline
Light cone sum rules \cite{2011} & $(3.03 \pm 0.10) \times 10^{-4} $ \\ 
 SU(3) flavor  symmetry \cite{SU(3)} & $(1.23 \pm 0.64)\times 10^{-5} $ \\ 
Light-cone QCD sum rules in full theory \cite{Rare} &
                                                      $1.08^{+0.63}_{-0.49} \times 10^{-5}$ \\ 
Light-front quark model \cite{lfqm} & $ (1.1 \pm 0.1)\times 10^{-5} $   \\  
This paper & $ (0.95 \pm 0.15)\times 10^{-5}$ \\ 
Experiment \cite {Experimental} & $<1.3 \times 10^{-4}$ \\ \hline
\end{tabular}
\end{center}
\end{table}

We compare our result with the previous theoretical predictions
\cite{2011,lfqm,SU(3),Rare} and the experimental upper limit in Table~\ref{tabular:timesandtenses}.
One can see, that the result of the light cone sum rules
Ref.~\cite{2011} is significantly higher than other theoretical
predictions and  it exceeds the experimental upper limit. Our result
is consistent with the values from Refs.~\cite{SU(3),lfqm, Rare} within theoretical uncertainties and these  values are lower than the experimental limit.  Thus, the measurement of the rare radiative $\Xi^{-}_{b} \rightarrow \Xi^{-}\gamma$ decay branching fractions can discriminate between different approaches.

\section{Conclusion}

The rare radiative decay $\Xi^{-}_{b} \rightarrow \Xi^{-}\gamma$ is
investigated in the framework of the relativistic quark model. First,
we give the expression for the effective Hamiltonian and evaluate the
relevant Wilson coefficient by solving the system of the
renormalization group equations. Then the quasipotential approach and
relativistic quark-diquark picture are employed for the calculation of
the form factors parameterizing the  hadronic matrix elements of the
weak current. The baryon wave functions are obtained with the account
of all relativistic effects including transformation from rest to the
moving reference frame and contributions of the intermediate
negative-energy states. The form factors of this decay are expressed
as the overlap integrals of the initial and final baryon wave
functions \cite{Galkin}. Using the calculated values of form factors
at $q^2=0$ (since the real photon is emitted) we obtained
the value of the branching ratio. Our result is consistent with the
predictions of the theoretical approaches in
Refs.~\cite{SU(3),lfqm,Rare} and is below the experimental upper limit
set by LHCb \cite{Experimental}, while the branching ratio obtained in
Ref.~\cite{2011} is significantly higher  and contradicts the
experimental limit. Therefore the exact experimental value is needed
to make a final comparison.

\acknowledgments
We are grateful to D. Ebert  for valuable
discussions.

\end{document}